\documentclass[aps, prd, amsmath, amssymb, showpacs, reprint]{revtex4-1}

\usepackage{mathrsfs} % \mathscr fonts (used for notating the spacetime manifold)
\usepackage[hyperindex, breaklinks, colorlinks=true, linkcolor={blue}, citecolor={red},urlcolor={blue}]{hyperref} % hyperrefs
\usepackage{tensor} % for indices
\newcommand{\ind}[1]{\indices{#1}} % redefining `indices' commands
\renewcommand{\d}{\text{d}} % differential
\newcommand{\R}{\ensuremath{\mathcal{R}}} % radial component of potential
\renewcommand{\L}{\ensuremath{\mathcal{L}}} % Lagrange density
\newcommand{\M}{\ensuremath{\mathscr{M}}} % manifold
\renewcommand{\vec}[1]{\ensuremath{\mathbf{#1}}} % bold vectors/matrices
\newcommand{\eref}[1]{(\ref{#1})} % equation reference
\newcommand{\sref}[1]{\text{Sec.}~\ref{#1}} % section reference
\newcommand{\pref}[1]{Eq.~(\ref{#1})} % equation reference (two)
\newcommand{\rref}[1]{Ref.~\cite{#1}} % reference reference
\newcommand{\rrefs}[2]{Refs.~\cite{#1, #2}} % reference reference (two)
\newcommand{\rrefss}[3]{Refs.~\cite{#1, #2, #3}} % reference reference (three)
\newcommand{\prefs}[2]{Eqs.~(\ref{#1})~and~(\ref{#2})} % equation reference (three)
\newcommand{\I}{\ensuremath{\vec{I}}} % identity matrix
\newcommand{\Tr}{\text{Tr}} % trace
\newcommand{\C}{\ensuremath{\mathcal{C}}} 
\renewcommand{\S}{\ensuremath{\mathcal{S}}}

\begin{document}

\title{Palatini $f(\R, \L_m, \R_{\mu\nu}T^{\mu\nu})$ gravity and its Born-Infeld semblance}

\author{Matthew S. Fox}

\email{msfox@g.hmc.edu}

\affiliation{Department of Physics, Harvey Mudd College, Claremont, Ca 91711, USA}

\date{\today}

\begin{abstract}
We investigate Palatini $f(\R,\L_m, \R_{\mu\nu}T^{\mu\nu})$ modified theories of gravity wherein the metric and affine connection are treated as independent dynamical fields and the gravitational Lagrangian is made a function of the Ricci scalar $\R$, the matter Lagrangian density $\L_m,$ and a ``matter-curvature scalar'' $\R_{\mu\nu}T^{\mu\nu}$. The field equations and the equations of motion for massive test particles are derived, and we show that the independent connection can be expressed as the Levi-Civita connection of an auxiliary, energy-momentum dependent metric, related to the physical metric by a matrix transformation. Similar to metric $f(\R, T, \R_{\mu\nu}T^{\mu\nu})$ gravity, the field equations impose the non-conservation of the energy-momentum tensor, leading to an appearance of an extra force on massive test particles. We obtain the explicit form of the field equations for massive test particles in the case of a perfect fluid, and an expression for the extra force. The nontrivial modifications to scalar fields and both linear and nonlinear electrodynamics are also considered. Finally, we detail the conditions under which the present theory is equivalent to the Eddington-inspired Born-Infeld (EiBI) model.
\end{abstract}


\maketitle

\section{Introduction} 

Observations of the cosmic microwave background (CMB) \cite{Netterfield, *Bennett, *Halverson} and direct measurements of the light curves from several hundred type Ia supernovae \cite{Riess, *Perlmutter, *Tonry} suggest that the Universe is presently undergoing a phase of late-time, accelerated expansion. While the physics underlying this phenomenon remain unsettled, at least one thing is certain: the acceleration is either a trait of the gravitational interaction itself, or it is the gravitational manifestation of something else (dark energy). By and large, the copious models of the former type derive from revisions to the Einstein-Hilbert action
\begin{equation}
\S_{\text{EH}} = \frac{1}{2\kappa_E}\int \d^4x\, \sqrt{-g}\R,
\label{eq:EinsteinHilbertAction}
\end{equation}
where $\kappa_E$ is the Einstein constant, $\R$ is the Ricci scalar, and $g$ is the determinant of the spacetime metric $g_{\mu\nu}$.

Perhaps the simplest set of modifications to $\S_{\text{EH}}$ are the $f(\R)$ models, which constitute a class of higher order gravity theories in which $\S_{\text{EH}}$ is restyled with terms of higher degree in the scalar curvature. Indeed, the mystery of cosmic expansion can be unraveled with this approach \cite{Starobinsky, *Carroll}. Although fatal instabilities and acute weak-field constraints appear to bar many of the proposals \cite{Stelle, *Woodard, *Chiba, *Erickcek, *Chiba2, *Capozziello, *Capozziello2, *Amarzguioui, *Koivisto, *Starobinsky2, *Bergliaffa, *Bohmer}, some persevere \cite{Hu, *Amendola, *Faraoni, *Faulkner, *Zhang, *Nojiri2, *Nojiri3, *Pun, *Sawicki}. Incidentally, the lure of $f(\R)$ is broader in application than to just cosmic speed-up. For instance, theories with higher order curvature invariants show promise as effective first-order approximations to quantum gravity, and can encourage quantum and gravitational fields to be well-behaved in the ultraviolet regions neighboring curvature singularities \cite{Vilkovisky, *Buchbinder, *Parker}. Further $f(\R)$ phenomenology has been extensively surveyed in the literature \cite{Felice, *Sotiriou}.

An interesting extension of the $f(\R)$ models are those which include in the action an explicit non-minimal coupling between matter and curvature invariants. One set of models in particular are the so-called $f(\R, \L_m)$ models ($\L_m$ being the matter Lagrangian density) originally proposed by  Bertolami \emph{et al.} \cite{Bertolami}. Their model was linear in the non-minimal coupling, which prompted \rref{Harko2} to study the maximal extension of $\S_{\text{EH}}$ in which $\R$ and $\L_m$ are coupled arbitrarily. Cosmological and astrophysical phenomena have been studied in the $f(\R,\L_m)$ framework \cite{Nojiri4, *Allemandi, *Bertolami3} in addition to more general studies into the properties of the theory itself \cite{Faraoni2, *Bertolami4, *Nesseris, *Puetzfeld}.

In general, non-minimal theories such as $f(\R,\L_m)$ gravity tend to rid one of the ability to locally transform away the influence of a gravitational field on matter \cite{Goenner, *Bahamonde}. In turn, the covariant divergence of the energy-momentum tensor is generally nonzero, the motion of test particles is generally non-geodesic (due to the presence of an extra force orthogonal to the four-velocity \cite{Bertolami}), and thus the equivalence principle (EP) is generally violated. Hence these theories are stringently constrained by tests of the EP. It is important to note, however, that a violation of the EP does not in principle disqualify the specific theory \cite{Bertolami2, *Damour, *Damour2}.

A set of models related to the $f(\R, \L_m)$ paradigm derives from the case in which the functional dependence on $\L_m$ manifests from a dependence on the trace $T$ of the energy-momentum tensor. These so-called $f(\R, T)$ models have drawn significant attention, and were explicitly introduced by Harko \emph{et al.} \cite{Harko, *Barrientos2}. However, Poplawski \cite{Poplawski} was first to consider a model in which the cosmological constant is a function of $T$, which is considered a relativistically covariant model for interacting dark energy, and which is evidently a subset of the $f(\R, T)$ paradigm. Related models have also been studied, notably \rref{Roshan}. We note that explicit dependences on $T$ may be induced by quantum effects (\emph{e.g.}, conformal anomalies) or exotic imperfect fluids. The reader is referred to the review \cite{Harko3} for additional $f(\R,\L_m)$ and $f(\R, T)$ phenomenology.

A further extension to the $f(\R,\L_m)$ and $f(\R,T)$ paradigms was proposed in \rrefs{Odintsov}{Haghani}, in which terms of the form $\R_{\mu\nu}T^{\mu\nu}$, where $\R_{\mu\nu}$ is the Ricci tensor, were incorporated into the $f(\R,T)$ Lagrangian. Instances of this coupling are known to arise in Born-Infeld models of gravity \cite{Deser} when one Taylor expands the Lagrangian. The cosmological implications of these so-called $f(\R, T, \R_{\mu\nu}T^{\mu\nu})$ gravity theories were surveyed in \rrefss{Haghani}{Odintsov}{Baffou, *Sharif2}, and the criterion to circumvent the Dolgov-Kawasaki instability \cite{Dolgov} can be found in \rref{Haghani}. Moreover, energy conditions and thermodynamic laws in $f(\R, T, \R_{\mu\nu}T^{\mu\nu})$ gravity were considered in \rref{Sharif}. Finally, it is known that metric $f(\R, T, \R_{\mu\nu}T^{\mu\nu})$ gravity acquires ghost-like instabilities due to the additional $\R_{\mu\nu}T^{\mu\nu}$ coupling \cite{Ayuso, *Ayuso2}, and that these instabilities can be avoided with a Palatini or metric-affine variation \cite{Afonso}.

The appearance of the $\R_{\mu\nu}T^{\mu\nu}$ coupling in Born-Infeld gravity is the chief motivation for the present study. The Born-Infeld models themselves, akin to Born-Infeld electromagnetism, modify the determinantal structure of $\S_{\text{EH}}$. Among these theories, a notable one is the Eddington-inspired Born-Infeld (EiBI) model proposed in \rref{Banados}. Many $f(\R)$ models will differ from GR even in vacuum; EiBI does not. Yet in ultraviolet regions, such as near cosmological singularities, EiBI gravity is characterized by curing the geometrical divergences plaguing GR \cite{Banados}. See \rref{Jimenez} for a recent review on Born-Infeld modifications to gravity.

Importantly, in all of these theories, independent of the details of the modification, one must ultimately choose between two ostensibly similar methods for varying the action: one either treats the metric as the sole dynamical entity and fixes \emph{a priori} the connection to be the Levi-Civita connection of $g_{\mu\nu}$ (the metric formalism) or one regards the metric and affine connections as independent dynamical structures (the metric-affine or Palatini formalisms, depending on whether matter couples to the connection or not, respectively). In GR the distinction is superfluous as they both lead to the same physics. However, in general, nearly all the aforementioned theories have differing physics depending on whether the metric and affine structures are treated independently or not. In fact, in some theories, such as EiBI gravity \cite{Banados} and, already mentioned, $f(\R,T,\R_{\mu\nu}T^{\mu\nu})$ gravity \cite{Afonso}, the Palatini formalism will remove ghost-like instabilities that otherwise afflict their metric counterparts. Whether the affine connection is determined by the metric degrees of freedom or not is a truly fundamental question and demands experimental investigation.

Though matter couplings to the connection may arise due to quantum gravitational corrections, we shall ignore that possibility here, and so we exercise the Palatini formalism. Studies of Palatini $f(\R)$ and $f(\R, T)$ models can be found in Refs.~\cite{Olmo2} and \cite{Barrientos, Wu}, respectively, and more general actions varied {\`a} la Palatini and metric-affine, including the role of torsion, can be found in \rref{Afonso}. To the best of our knowledge, no studies of pure Palatini $f(\R, T, \R_{\mu\nu}T^{\mu\nu})$ or Palatini $f(\R, \L_m, \R_{\mu\nu}T^{\mu\nu})$ gravity have yet been completed, though indirect pursuits exist (see, \emph{e.g.}, \rref{Afonso}). In this paper, we shall investigate Palatini $f(\R, \L_m, \R_{\mu\nu}T^{\mu\nu})$ gravity, from which Palatini $f(\R, T, \R_{\mu\nu}T^{\mu\nu})$ gravity follows after a simple modification to the field equations. In addition to studying $f(\R,\L_m, \R_{\mu\nu}T^{\mu\nu})$ gravity on its own, we ultimately seek the conditions under which our theory corresponds to EiBI gravity.

The present paper is structured as follows. In \sref{sec:EOM}, we vary the $f(\R, \L_m, \R_{\mu\nu}T^{\mu\nu})$ action {\`a} la Palatini and derive the theory's equations of motion and an explicit form for the independent connection. In \sref{sec:commentsOnFE}, we survey the bimetric structure of $f(\R,\L_m, V)$ gravity in addition to the non-minimal structure of the field equations. In \sref{sec:Properties}, we explore various properties of the $f(\R, \L_m, \R_{\mu\nu}T^{\mu\nu})$ field equations, including their non-conservation equation, the non-geodesic motion of test particles, the nature of the extra force, the weak-field limit, and the modified Poisson equation. In \sref{sec:applications}, we derive the $f(\R,\L_m, \R_{\mu\nu}T^{\mu\nu})$ field equations for the cases of linear and nonlinear electromagnetic fields, as well as canonical scalar fields. Finally, in \sref{sec:EiBI} we derive the conditions under which the $f(\R, \L_m, \R_{\mu\nu}T^{\mu\nu})$ model responds identically to the EiBI theory for specific matter sectors.

In this paper we shall operate in a four-dimensional spacetime $(\M, g_{\mu\nu}, \Gamma_{\mu\nu}^\alpha)$ in which the metric $g_{\mu\nu}$ and connection $\Gamma_{\mu\nu}^\alpha$ are assumed to be independent dynamical fields. We shall utilize the metric signature $(-,+,+,+)$ and, where appropriate, adopt the natural system of units in which $c = 8\pi G = 1$.

\section{Field equations of $f(\R, \L_m, \R_{\mu\nu}T^{\mu\nu})$ gravity}
\label{sec:EOM}

The Ricci tensor can be defined solely in terms of the affine connection, and this underpins the Palatini and metric-affine formalisms. Explicitly, the Ricci tensor follows from the Riemann curvature tensor
\begin{equation}
\R\ind{^\alpha_\beta_\mu_\nu} = \partial_\mu \Gamma^\alpha_{\nu\beta} - \partial_\nu\Gamma_{\mu\beta}^\alpha + \Gamma_{\mu\lambda}^\alpha\Gamma^\lambda_{\nu\beta} - \Gamma_{\nu\lambda}^{\alpha}\Gamma_{\mu\beta}^{\lambda}
\label{eq:RiemannTensor}
\end{equation}
via the contraction $\R_{\mu\nu}(\Gamma) \equiv \R\ind{^\alpha_\mu_\alpha_\nu}(\Gamma)$. Only now need one invoke the metric to define the Ricci scalar $\R(g, \Gamma) \equiv g^{\mu\nu}\R_{\mu\nu}(\Gamma)$ and a \emph{matter-curvature scalar} $V(g, \Gamma, \Psi) \equiv \R_{\mu\nu}(\Gamma)T^{\mu\nu}(g,\Psi)$, where $T_{\mu\nu}$ is the symmetric (Hilbert) energy-momentum tensor. As we shall see below [\pref{eq:EnergyMomentumDefinition}], the energy-momentum tensor is constructed {\`a} la Palatini so that $T_{\mu\nu}$ depends only on the metric and a set of matter fields $\Psi$. We note that the symmetry of $g_{\mu\nu}$ and $T_{\mu\nu}$ impose that only the symmetric part of the Ricci tensor enters into this theory's action. This considerably simplifies the role of torsion in the theory, and renders a separate consideration for fermionic matter immaterial \cite{Afonso}. 

With all this in mind, the action considered in this work bears the form
\begin{equation}
\S[g, \Gamma, \Psi] = \frac{1}{2\kappa}\int \d^4x\; \sqrt{-g}f(\R,\L_m, V) + \S_m[g, \Psi],
\label{eq:f(R,V)Action}
\end{equation}
where $\kappa$ is a coupling constant with suitable dimensions. Here the matter Lagrangian density $\L_m$, encoded in both the function $f(\R, \L_m, V) = f(\R,\L_m, \R_{\mu\nu}T^{\mu\nu})$ and the matter action
\begin{equation}
\S_m[g,\Psi] = \int \d^4x\, \sqrt{-g}\L_m[g, \Psi],
\label{eq:MatterSectorAction}
\end{equation} 
is assumed to capture all matter fields $\Psi$ present in $\M$. Moreover, $\L_m$ determines the manifestly symmetric energy-momentum tensor
\begin{equation}
T_{\mu\nu} \equiv -\frac{2}{\sqrt{-g}} \frac{\delta\left(\sqrt{-g}\L_m\right)}{\delta g^{\mu\nu}},
\label{eq:EnergyMomentumDefinition}
\end{equation}
which again is independent of the affine connection in the Palatini formulation.

If we denote by $\delta \S_g$ and $\delta \S_\Gamma$ the variation of \pref{eq:f(R,V)Action} with respect to the metric and connection, respectively, then $\delta \S = \delta \S_g + \delta \S_\Gamma$ with
\begin{multline}
\delta \S_g = \frac{1}{2\kappa} \int \d^4x\, \sqrt{-g}\bigg[ -\frac{1}{2}fg_{\mu\nu} + f_{\R}\R_{\mu\nu}\\ + f_{\L}\Xi_{\mu\nu} + f_V \Pi_{\mu\nu} + \frac{1}{\sqrt{-g}}\frac{\delta(\sqrt{-g}\L_m)}{\delta g^{\mu\nu}}\bigg]\delta g^{\mu\nu}
\label{eq:variationMetric}
\end{multline}
and
\begin{equation}
\delta \S_\Gamma = \frac{1}{2\kappa}\int \d^4x\, \sqrt{-g}\left[\left(f_\R g^{\mu\nu} + f_V T^{\mu\nu}\right)\frac{\delta \R_{\mu\nu}}{\delta \Gamma_{\alpha\beta}^\lambda}\right]\delta \Gamma_{\alpha\beta}^\lambda.
\label{eq:variationConnection}
\end{equation}
Here we have introduced the definitions: $f_{\R} \equiv \partial_\R f$, $f_V \equiv \partial_V f, f_{\L} \equiv \partial_{\L_m} f$, as well as the manifestly symmetric \emph{matter} and \emph{matter-curvature tensors}
\begin{align}
\Xi_{\mu\nu} &\equiv \frac{\partial \L_m}{\partial g^{\mu\nu}},
\label{eq:XiTensor}\\
\Pi_{\mu\nu} &\equiv \R^{\alpha\beta} \frac{\delta T_{\alpha\beta}}{\delta g^{\mu\nu}},
\label{eq:PiTensor}
\end{align}
respectively. Since $g_{\mu\nu}$ and $\Gamma_{\mu\nu}^\alpha$ are assumed to be independent of each other, $\delta \S = 0$ if and only if $\delta \S_g$ and $\delta \S_\Gamma$ vanish separately. In the case of the metric variation \eref{eq:variationMetric}, $\delta \S_g = 0$ and \pref{eq:EnergyMomentumDefinition} imply
\begin{equation}
f_\R \R_{\mu\nu} - \frac{1}{2}fg_{\mu\nu} = \kappa T_{\mu\nu} - f_{\L}\Xi_{\mu\nu} - f_V\Pi_{\mu\nu}.
\label{eq:FieldEquationOne}
\end{equation}
This is the $f(\R, \L_m, V)$ generalization of Einstein's equation. Its properties shall be explored in the coming sections. We note here, however, that as a consequence of the non-minimal coupling, there appears in \pref{eq:FieldEquationOne} strict couplings between matter fields and curvature terms. This is very much unlike GR and other minimally coupled theories in which matter fields are wholly separable from curvature terms such that the field equations may be written in a ``curvature = matter'' type representation. Ultimately, however, writing the field equations in this way is more for physical tidiness and less for mathematical substance. Hence the mathematical representation of these equations may as well be chosen such that it facilitates later computation. To this end, we define a curvature-dependent \emph{effective} energy-momentum tensor by
\begin{equation}
\Sigma_{\mu\nu} \equiv T_{\mu\nu} - \frac{f_{\L}}{\kappa}\Xi_{\mu\nu} - \frac{f_V}{\kappa} \Pi_{\mu\nu},
\label{eq:SigmaTensor}
\end{equation}
which refashions the field equations \eref{eq:FieldEquationOne} into a form similar to those in Palatini $f(\R)$ gravity:
\begin{equation}
f_\R\R_{\mu\nu} - \frac{1}{2}fg_{\mu\nu} = \kappa \Sigma_{\mu\nu}.
\label{eq:FieldEquationOneTwo}
\end{equation}

The variation with respect to the connection takes more care. We refer the reader to \rref{Afonso} in which a nearly complete derivation is given. One shall find that $\delta \S_\Gamma = 0$ only if\begin{equation}
\nabla^{(p)}_\sigma \left[\sqrt{-g}\left(f_\R g^{\mu\nu} + f_V T^{\mu\nu}\right)\right] = 0,
\label{eq:FieldEquationTwo}
\end{equation}
where $\nabla^{(p)}$ is the derivative operator associated with the independent connection, and which is manifestly distinct from $\nabla^{(g)}$, the covariant derivative compatible with the spacetime metric $g_{\mu\nu}$. The resemblance of \pref{eq:FieldEquationTwo} to the companion EiBI field equation will be studied in \sref{sec:EiBI}. 

We note that \pref{eq:FieldEquationTwo} holds good even in the presence of torsion. This follows from this theory's insensitivity to the projective degrees of freedom in projective transformations of the independent connection, which ultimately derives from only the symmetric part of the Ricci tensor entering into the action \eref{eq:f(R,V)Action}. See \rref{Afonso} for details.

Together \prefs{eq:FieldEquationOne}{eq:FieldEquationTwo} comprise the field equations of $f(\R, \L_m, V)$ gravity. We see in \pref{eq:FieldEquationTwo} a natural auxiliary metric ingrained into this theory's mathematical structure, namely a metric $p_{\mu\nu}$ whose \emph{inverse}, denoted $p^{\mu\nu}$,\footnote{We emphasize: $p^{\mu\nu}$ is defined to satisfy $p^{\mu\lambda}p_{\lambda\nu} = \delta\ind{^\mu_\nu}$ and is in general different from $g^{\mu\alpha}g^{\nu\beta}p_{\alpha\beta}$.} satisfies
\begin{equation}
\sqrt{-p}p^{\mu\nu} = \sqrt{-g}\left(f_\R g^{\mu\nu} + f_VT^{\mu\nu}\right),
\label{eq:pmunuDefinition}
\end{equation}
where $p \equiv \det(p_{\mu\nu})$. Evidently, the symmetry of $g^{\mu\nu}$ and $T^{\mu\nu}$ force $p^{\mu\nu}$, and hence $p_{\mu\nu}$, to be symmetric. Moreover, $p^{\mu\nu}$ satisfies $\nabla^{(p)}_\sigma (\sqrt{-p}p^{\mu\nu}) = 0$ by construction, and hence $p_{\mu\nu}$ is compatible with $\nabla^{(p)}$ provided the coefficients of the independent connection are the Christoffel symbols in $p_{\mu\nu}$,
\begin{equation}
\Gamma_{\mu\nu}^\alpha = \frac{1}{2}p^{\alpha\sigma}\left(\partial_{\mu}p_{\sigma\nu} + \partial_{\nu}p_{\mu\sigma} - \partial_\sigma p_{\mu\nu}\right).
\label{eq:connectionEquation}
\end{equation}
Consequently, the independent connection is the Levi-Civita connection in the auxiliary metric $p_{\mu\nu}$. Note also that the determinant $p$ can be computed explicitly with \pref{eq:pmunuDefinition} and the relation $p = \det^{-1}(p^{\mu\nu})$. One finds
\begin{equation}
p = g^2\det\left(f_\R g^{\mu\nu} + f_V T^{\mu\nu}\right).
\label{eq:deteriminantFormula}
\end{equation}
We shall apply these formulae to various physical phenomena in the coming sections. But first we briefly comment on some general characteristics of the field equations.

\section{Remarks on the $f(\R,\L_m, V)$ field equations}
\label{sec:commentsOnFE}

As noted above, for theories in which couplings are minimal, the matter fields can in general be placed on one side of the theory's field equation, and the symmetric part of the Ricci tensor will be given solely in terms of $g_{\mu\nu}$. But for non-minimal theories, the matter fields are generally inseparable from the geometry terms, and the symmetric part of the Ricci tensor need not be given solely in terms of $g_{\mu\nu}$. Such is the case for $f(\R, \L_m, V)$ gravity, as made evident by the field equations \eref{eq:FieldEquationOne} and \eref{eq:FieldEquationTwo}. Other aspects of the present theory's non-minimal character are addressed in this section. 

\subsection{The matter and matter-curvature tensors}

The matter-curvature tensor $\Pi_{\mu\nu}$ is a hallmark of the present theory's non-minimal coupling. For the sake of computation, it is of interest to write this tensor in a form entirely in terms of the matter Lagrangian and Ricci tensor. To this end, assuming the matter Lagrangian is independent of derivatives of the metric, one can show that \pref{eq:EnergyMomentumDefinition} is equivalent to
\begin{equation}
T^{\mu\nu} = \L_m g^{\mu\nu} + 2 \frac{\partial \L_m}{\partial g_{\mu\nu}}.
\label{eq:EMTensorTwo}
\end{equation}
Incidentally, this equation has the matter tensor $\Xi_{\mu\nu}$ implicitly built into it,
\begin{equation}
\Xi_{\mu\nu} = \frac{1}{2}\left(\L_m g_{\mu\nu} - T_{\mu\nu}\right),
\label{eq:XiAlternative}
\end{equation} 
which we shall find useful later on. Moreover, \pref{eq:EMTensorTwo} facilitates calculating the functional derivative
\begin{equation}
\frac{\delta T^{\alpha\beta}}{\delta g^{\mu\nu}} = g^{\alpha\beta}\frac{\partial \L_m}{\partial g^{\mu\nu}} + 2\frac{\partial^2\L_m}{\partial g^{\mu\nu}\partial g_{\alpha\beta}} + \L_m \delta\ind{^(^\alpha^\beta^)_\mu_\nu},
\label{eq:VariationSquared}
\end{equation}
where $\delta\ind{^(^\alpha^\beta^)_\mu_\nu} = \frac{1}{2}\left(\delta\ind{^\alpha_\mu} \delta\ind{^\beta_\nu} + \delta\ind{^\beta_\mu} \delta\ind{^\alpha_\nu}\right)$ is the upper symmetric part of the generalized Kronecker symbol (we herein denote symmetrization by parentheses). It then follows from \pref{eq:VariationSquared}, the definition of the matter-curvature tensor in \eref{eq:PiTensor}, and the fact that $\R_{\mu\nu}$ is the symmetric part of the Ricci tensor that
\begin{equation}
\Pi_{\mu\nu} = \R\frac{\partial \L_m}{\partial g^{\mu\nu}} + 2\R_{\alpha\beta}\frac{\partial^2\L_m}{\partial g^{\mu\nu}\partial g_{\alpha\beta}} + \R_{\mu\nu}\L_m.
\label{eq:PiV1}
\end{equation}
Another useful identity is the following:
\begin{equation}
\Pi_{\mu\nu} = 2\R_{\lambda(\mu}T\ind{^\lambda_{\nu)}} + \R^{\alpha\beta}\frac{\delta T_{\alpha\beta}}{\delta g^{\mu\nu}},
\label{eq:PiV2}
\end{equation}
which follows from substituting $T^{\mu\nu} = g^{\mu\alpha}g^{\nu\beta}T_{\alpha\beta}$ into the definition \eref{eq:PiTensor} of the matter-curvature tensor.

A notable matter sector is that of a perfect fluid (PF), for which we shall take $\L_m = P$,\footnote{See, \emph{e.g.}, \rrefss{Harko4}{Bertolami5}{Faraoni2} for discussions on the issue of whether $\L_m = -\rho$ or $\L_m = P$ is the correct Lagrangian for a perfect fluid, and the consequences in non-minimally coupled theories.} where $P$ is the isotropic pressure of the fluid. The corresponding energy-momentum tensor is
\begin{equation}
T_{\mu\nu}^{(\text{PF})} = (\rho + P)u_\mu u_\nu + P g_{\mu\nu},
\label{eq:StressEnergyPerfectFluid}
\end{equation}
where $\rho$ is the energy density of the fluid and the fluid's four-velocity $u^\mu$ satisfies the condition $u_\mu u^\mu = -1$. One can show the pressure $P$ satisfies 
\begin{equation}
\delta P = -\frac{1}{2}\left(\rho + P\right)u_\mu u_\nu \delta g^{\mu\nu}
\label{eq:}
\end{equation}
by using \pref{eq:EMTensorTwo} with $\L_m = P$ and comparing to \pref{eq:StressEnergyPerfectFluid}. Moreover, one has \cite{Harko4}
\begin{equation}
\delta \rho = \frac{1}{2}\rho(g_{\mu\nu} - u_\mu u_\nu)\delta g^{\mu\nu},
\label{eq:dustPerturbation}
\end{equation}
which ultimately follows from the conservation of the matter fluid current, $\nabla_\mu^{(g)} (\rho u^\mu) = 0$. Using these formulae appropriately, one shall find 
\begin{equation}
\Xi_{\mu\nu}^{(\text{PF})} = -\frac{1}{2}\left(\rho + P\right)u_\mu u_\nu
\label{eq:XiPerfectFluid}
\end{equation}
and, from \pref{eq:PiV2} and the identity $\frac{\delta u_\alpha}{\delta g^{\mu\nu}} = -\frac{1}{2}g_{\alpha(\mu}u_{\nu)}$, 
\begin{multline}
\Pi_{\mu\nu}^{(\text{PF})} = \R\ind{^\lambda_(_\mu}T_{\nu)\lambda}^{(\text{PF})} + \frac{1}{2}\rho\R^{\alpha\beta}u_\alpha u_\beta g_{\mu\nu}\\ - \frac{1}{2}\left[\left(2\rho + P\right) \R^{\alpha\beta}u_\alpha u_\beta + (\rho + P) \R \right]u_\mu u_\nu.
\label{eq:PiPerfectFluid}
\end{multline}
The effective energy-momentum tensor for a perfect fluid then follows from its definition \eref{eq:SigmaTensor}. The rather exotic coupling of matter and the four-velocity to the Ricci tensor in \pref{eq:PiPerfectFluid} suggests that the matter-curvature tensor will play a significant role in the field equations in regions of high density, such as within a black hole or in the very early universe. This is quantitatively similar to EiBI gravity, which has in its field equations a similar $\R_{\mu\nu}T^{\mu\nu}$ coupling that too gives rise to couplings between the Ricci tensor and the four-velocity of perfect fluids (see \rref{Jimenez} or \sref{sec:EiBI} of this paper). It is natural to hypothesize, then, that $f(\R, \L_m, V)$ gravity may be fashionable such that it corresponds to GR in the weak-field regime but then cures the curvature singularities of GR in high density regions---behavior that mimics the preeminent characteristics of EiBI gravity.

\subsection{The auxiliary metric}

The introduction of the ``natural'' auxiliary metric $p_{\mu\nu}$ into the present theory affords a certain bimetric nature to the $f(\R,\L_m, V)$ model: in addition to the physical spacetime metric $g_{\mu\nu},$ through which the gravitational observables manifest, there is the auxiliary metric upon which the mathematical edifice of the theory is most well supported. This structure is analogous to EiBI gravity wherein there too exists a natural bimetric construction \cite{Banados, Jimenez}. In the present theory, however, unlike the minimal nature of EiBI theory, there is built explicitly into the gravitational Lagrangian an indirect coupling between the matter fields and the auxiliary metric via the scalar curvature $\R$ and the matter-curvature scalar $V \equiv \R_{\mu\nu} T^{\mu\nu}$. This appears through the explicit dependence of the independent connection \eref{eq:connectionEquation} on $p_{\mu\nu}$ and its inverse. This coupling suggests there exists some physicality associated with the auxiliary metric, which necessarily manifests via the spacetime metric. While the particulars of the physical meaning cannot be properly realized until the specific nature of the non-minimal coupling is known (which necessitates specifying a particular function $f$), it suggests a general link between the two metrics. A standard position is that of a conformal relationship, in which $p_{\mu\nu} = \Theta^2 g_{\mu\nu}$ for some real-valued, smooth function $\Theta$ defined on $\M$. This approach, however, is consistent only for specific matter distributions.\footnote{To see this, first use the spacetime dimensionality condition $p\ind{^\mu_\mu} = 4$ to derive an exact expression for $\Theta^2$ in terms of $f_\R$, $f_V$, and $T\ind{^\mu_\mu}$. Then one shall find, for instance, that the reciprocal identity $p^{\mu\lambda}p_{\lambda\nu} = \delta\ind{^\mu_\nu}$ holds only for particular energy-momentum tensors.} Thus, conformality between $p_{\mu\nu}$ and $g_{\mu\nu}$ fails to capture the general framework we seek. A more general approach, again analogous to EiBI gravity, is to introduce a differentiable \emph{deformation matrix} $\Omega\ind{^\mu_\nu}$ satisfying
\begin{equation}
p_{\mu\nu} = g_{\mu\lambda}\Omega\ind{^\lambda_\nu}.
\label{eq:deformationMatrix}
\end{equation}
In matrix notation this reads $\vec{p} = \vec{g}\vec{\Omega}$ so that $\vec{p}^{-1} = \vec{\Omega}^{-1}\vec{g}^{-1}$. Direct comparison to \pref{eq:pmunuDefinition} reveals that
\begin{equation}
\vec{\Omega}^{-1} = \frac{1}{\sqrt{\Omega}}\left(f_\R \I + f_V \vec{g}^{-1}\vec{T}\right),
\label{eq:OmegaMatrix}
\end{equation}
where $\Omega = \det(\vec{\Omega})$ [whose value follows from \pref{eq:deteriminantFormula}] and $\I$ is the identity matrix. It is now an algebra problem to solve for $\vec{\Omega}$, and hence $p_{\mu\nu}$, explicitly, following the specification of the matter Lagrangian and the $f(\R,\L_m, V)$ model of interest. One subsequently obtains the form of the connection and related curvature terms for the specific theory, and all that remains to resolve a given problem are the differential equations \eref{eq:FieldEquationOne}. An example of this procedure, in the context of EiBI gravity, can be found in \rref{Jimenez}.

\subsection{Likeness to other $f$ theories}

The Palatini $f(\R, \L_m, V)$ formalism surveyed here is the superset theory containing as special cases the Palatini $f(\R)$ and $f(\R, \L_m)$ theories, but not in general the Palatini $f(\R, T)$ and $f(\R, T, V)$ theories. Evidently, Palatini $f(\R, \L_m, V)$ and $f(\R, T, V)$ gravity correspond only when $\L_m = T$, which is a hefty constraint to which most matter fields do not abide.\footnote{However, it is argued in \rrefs{Avelino}{Avelino2} that the average on-shell Lagrangian for a soliton-constituted perfect fluid obeys this constraint. In this context, $f(\R, T, V)$ gravity may be considered a superset of $f(\R, \L_m, V)$ gravity.} That said, for matter fields with a vanishing energy-momentum trace (such as electromagnetic fields), the $f(\R, \L_m, V)$ paradigm clearly envelops the $f(\R, T, V)$ model. We say that Palatini $f(\R, \L_m, V)$ and $f(\R, T, V)$ are \emph{circumstantially equivalent} theories of gravity since their equivalence is such that it holds only for specific matter fields (this notion is made more precise in \sref{sec:EiBI}). There is, however, a simple procedure to obtain Palatini $f(\R, T, V)$ gravity from the $f(\R,\L_m, V)$ paradigm for arbitrary matter sectors: merely replace the $f_\L \Xi_{\mu\nu}$ term in the field equations \eref{eq:FieldEquationOne} by $f_T\frac{\partial T}{\partial g^{\mu\nu}}$, and continue on that way. Since in the Palatini formalism the trace $T \equiv T\ind{^\mu_\mu}$ is independent of the independent connection, its incorporation into the function $f$ will not affect \pref{eq:FieldEquationTwo}. In this respect, most results derived herein afford similar mathematical structure to Palatini $f(\R, T, V)$ gravity, up to the replacement of all $f_\L\Xi_{\mu\nu}$ terms with $f_T \frac{\partial T}{\partial g^{\mu\nu}}$ terms and the subsequent manipulations of those terms. Evidently, the exception to this rule are those results which utilize, in a nontrivial manner, the full entourage of dependencies in the $f(\R,\L_m, V)$ model, such as the present theory's circumstantial equivalence to EiBI gravity, which is discussed in detail in \sref{sec:EiBI}.

\section{Properties of the $f(\R, \L_m, \R_{\mu\nu}T^{\mu\nu})$ field equations}
\label{sec:Properties}

Here we shall consider various properties of the $f(\R,\L_m, V)$ field equations, including their conservation equation, their effect on the motion of massive test particles, and their weak-field limit.
\subsection{Conservation equation}
\label{ssref:conservation}

In $f(\R,\L_m, V)$ gravity, matter is non-minimally coupled to curvature. Hence the covariant divergence of the energy-momentum tensor is not necessarily zero. In this section we derive an explicit expression for such non-conservation of the energy-momentum tensor. In what follows, we decorate with tildes those tensors which have been raised/lowered by the auxiliary metric $p_{\mu\nu}$.

We start by writing the field equations \eref{eq:FieldEquationOneTwo} in the form
\begin{equation}
\widetilde{G}\ind{^\mu_\nu} = \frac{1}{f_\R}\left[\kappa\widetilde{\Sigma}\ind{^\mu_\nu} + \frac{1}{2}f(\Omega^{-1})\ind{^\mu_\nu}\right] - \frac{1}{2}\delta\ind{^\mu_\nu}\widetilde{\R},
\label{eq:VTwoFieldEquation1}
\end{equation}
where $\widetilde{G}\ind{^\mu_\nu}$ is the Einstein tensor raised by $p^{\mu\nu}$ and $(\Omega^{-1})\ind{^\mu_\nu}$ is the index equivalent of \pref{eq:OmegaMatrix}. Note that the definition \eref{eq:pmunuDefinition} provides an explicit form for the tensors $\widetilde{\Sigma}\ind{^\mu_\nu}$ and $\widetilde{\R}$:
\begin{subequations}
\label{eq:tensorFormulas}
\begin{align}
\widetilde{\Sigma}\ind{^\mu_\nu} &= \sqrt{\frac{g}{p}}\left(f_\R\Sigma\ind{^\mu_\nu} + f_V T^{\mu\lambda}\Sigma_{\lambda\nu}\right),\label{eq:SigmaTilde}\\
\widetilde{\R} &= \sqrt{\frac{g}{p}}\left(\R f_\R + V f_V\right).\label{eq:RicciTilde}
\end{align}
\end{subequations}
The condition we seek shall follow from $\nabla^{(p)}_{\mu}\widetilde{G}\ind{^\mu_\nu} = 0$, a consequence of Bianchi's identities. It remains an algebraic task to expand $\Sigma\ind{^\mu_\nu}$ in \pref{eq:SigmaTilde} and isolate the divergence of $T\ind{^\mu_\nu}$. We find
\begin{widetext}
\begin{multline}
\kappa\nabla_\mu^{(p)} T\ind{^\mu_\nu} = \sqrt{\frac{p}{g}}\bigg\{\nabla_{\mu}^{(p)} \left[\sqrt{\frac{g}{p}}\left(f_\L \Xi\ind{^\mu_\nu} + f_V \Pi\ind{^\mu_\nu} - \frac{f_V}{f_\R}T^{\mu\lambda}\Sigma_{\lambda\nu} - \frac{ff_V}{2f_\R}T\ind{^\mu_\nu}\right)\right] \\- \frac{1}{2}\partial_\nu\left(\sqrt{\frac{g}{p}}\left(f-\R f_\R- Vf_V\right)\right) - \kappa T\ind{^\mu_\nu}\partial_\mu \left(\sqrt{\frac{g}{p}}\right)\bigg\}.
\label{eq:nonConservationTwo}
\end{multline}
\end{widetext}
Alternatively, this non-conservation can be expressed in terms of the connection $\nabla^{(g)}$ compatible with the spacetime metric $g_{\mu\nu}$. The relationship between the covariant derivatives $\nabla^{(p)}$ (that defined with the independent connection of the auxiliary metric) and $\nabla^{(g)}$  is the following:
\begin{equation}
\nabla_\mu^{(p)} T\ind{^\mu_\nu} = \nabla_\mu^{(g)} T\ind{^\mu_\nu} + \C^\mu_{\mu\lambda}T\ind{^\lambda_\nu} - \C^\lambda_{\mu\nu}T\ind{^\mu_\lambda},
\label{eq:CovariantRelationship}
\end{equation}
where
\begin{equation}
\C_{\mu\nu}^\alpha = \frac{1}{2}p^{\alpha\sigma}\left(\nabla^{(g)}_\mu p_{\sigma\nu} + \nabla^{(g)}_\nu p_{\mu\sigma} - \nabla^{(g)}_\sigma p_{\mu\nu}\right).
\label{eq:AuxiliaryConnection}
\end{equation}
Using the metric/auxiliary metric relationship \eref{eq:deformationMatrix}, the compatibility of $g_{\mu\nu}$ with $\nabla^{(g)}$, and the symmetry property of the auxiliary metric, these coefficients manipulate into a form independent of $p_{\mu\nu}$:
\begin{multline}
\C^\alpha_{\mu\nu} = \frac{1}{2}(\Omega^{-1})\ind{^\alpha_\sigma}\left(\nabla^{(g)}_\mu \Omega\ind{^\sigma_\nu} + \nabla^{(g)}_\nu \Omega\ind{^\sigma_\mu}\right) \\ - \frac{1}{2}g_{\mu \lambda}(\Omega^{-1})\ind{^\alpha_\sigma} \nabla^{\sigma}_{(g)} \Omega\ind{^\lambda_\nu}.
\label{eq:coefficientsOne}
\end{multline}
We note that any covariant derivative with respect to $\Omega\ind{^\mu_\nu}$ can be replaced by a derivative with respect to $(\Omega^{-1})\ind{^\mu_\nu}$ as their inverse relationship implies
\begin{equation}
(\Omega^{-1})\ind{^\lambda_\nu}\nabla_\sigma^{(g)} \Omega\ind{^\mu_\lambda} + \Omega\ind{^\mu_\lambda}\nabla_\sigma^{(g)} (\Omega^{-1})\ind{^\lambda_\nu} = 0.
\label{eq:inversionRelation}
\end{equation}
With this in mind, the coefficients \eref{eq:coefficientsOne} become
\begin{multline}
\C^\alpha_{\mu\nu} = -\frac{1}{2} \Omega\ind{^\lambda_\nu}\nabla^{(g)}_\mu(\Omega^{-1})\ind{^\alpha_\lambda} - \frac{1}{2} \Omega\ind{^\lambda_\mu}\nabla^{(g)}_\nu(\Omega^{-1})\ind{^\alpha_\lambda}\\ + \frac{1}{2}g_{\mu\lambda} (\Omega^{-1})\ind{^\alpha_\sigma}\Omega\ind{^\gamma_\nu}\Omega\ind{^\lambda_\epsilon}\nabla_{(g)}^\sigma (\Omega^{-1})\ind{^\epsilon_\gamma}
\label{eq:ConnectionDeformation}
\end{multline}
and the non-conservation of the energy-momentum tensor turns out to be
\begin{widetext}
\begin{multline}
\nabla_\mu^{(g)} T\ind{^\mu_\nu} = \frac{1}{\kappa}\sqrt{\frac{p}{g}}\bigg\{\nabla_\mu^{(p)}\left[\sqrt{\frac{g}{p}}\left(f_\L \Xi\ind{^\mu_\nu} + f_V \Pi\ind{^\mu_\nu} - \frac{f_V}{f_\R}T^{\mu\lambda}\Sigma_{\lambda\nu} - \frac{ff_V}{2f_\R}T\ind{^\mu_\nu}\right)\right]\\ - \frac{1}{2}\partial_\nu\left(\sqrt{\frac{g}{p}}\left[f-\R f_\R- Vf_V\right]\right)\bigg\} + \C^\lambda_{\mu\nu}T\ind{^\mu_\lambda}.
\label{eq:nonconservationTwo}
\end{multline}
\end{widetext}
Unfortunately, by our tests, the curvature and energy-momentum dependences in \pref{eq:nonconservationTwo} and the energy-momentum dependence of $(\Omega^{-1})\ind{^\mu_\nu}$ restrict these formulae from simplifying much beyond what is given here. We emphasize, therefore, that \pref{eq:nonconservationTwo} does not in general vanish. Hence the energy-momentum tensor in Palatini $f(\R,\L_m, V)$ gravity is in general not conserved. On the other hand, in \sref{sec:EiBI} we shall indirectly derive two nontrivial functions of $f(\R,\L_m, V)$ for which the covariant divergence of specific but nontrivial $T_{\mu\nu}$ necessarily vanish, hence conserving the energy-momentum tensor. That said, this conservation will not be obvious at the level of \pref{eq:nonconservationTwo}, though it will nevertheless be true. Finally, we note that for the Einstein-Hilbert model $f(\R, \L_m, V) = \R - 2\Lambda$, the conservation of $T_{\mu\nu}$ is restored, as desired.

\subsection{Motion of test particles}

For clarity, we denote by $\Delta_\nu$ the righthand side of \pref{eq:nonconservationTwo}. Then $\nabla_\mu^{(g)}T\ind{^\mu_\nu} = \Delta_\nu$. For the case of a perfect fluid, for which $T_{\mu\nu} = (\rho + P)u_\mu u_\nu + P g_{\mu\nu}$, it is straightforward to show, using the constraint from the conservation of the matter  fluid current, $\nabla_\mu^{(g)}(\rho u^\mu) = 0$, that 
\begin{equation}
u^\mu\nabla_\mu^{(g)} u^\nu = \frac{\Delta^\nu - u^\nu \nabla^{(g)}_\mu(P u^\mu) - \partial^\nu P}{P + \rho}.
\label{eq:perfectGeodesicEquation}
\end{equation}
Here the lefthand side coincides with the well known identity
\begin{equation}
u^\mu\nabla^{(g)}_\mu u^\nu = \frac{d^2x^\nu}{ds^2} + \Gamma^\nu_{\alpha\beta}\frac{dx^\alpha}{ds}\frac{dx^\beta}{ds}.
\label{eq:GeodesicOne}
\end{equation}
Therefore, \pref{eq:perfectGeodesicEquation} is the equation of motion for particles in the presence of an isotropic pressure $P$. Absent this pressure, the equation reduces to
\begin{equation}
\frac{d^2x^\nu}{ds^2} + \Gamma^\nu_{\alpha\beta}\frac{dx^\alpha}{ds}\frac{dx^\beta}{ds} = f^\nu,
\label{eq:EOMDustOne}
\end{equation}
where the extra force $f^\nu = \rho^{-1}\Delta^\nu_{(P = 0)}$ with
\begin{widetext}
\begin{multline}
\Delta_\nu^{(P = 0)} = \frac{1}{\kappa}\sqrt{\frac{p}{g}}\bigg\{\nabla_\mu^{(p)}\bigg[-\sqrt{\frac{g}{p}}\frac{\rho}{2}f_\L u^\mu u_\nu + \frac{f_V\rho}{2}\sqrt{\frac{g}{p}}\left(\left[\R\ind{^\sigma^\mu}u_{\nu} + \R\ind{^\sigma_\nu}u^\mu\right]u_\sigma + \R^{\alpha\beta}u_{\alpha}u_{\beta} \left[\delta\ind{^\mu_\nu} - 2u^\mu u_\nu\right] - \R u^{\mu}u_\nu\right)\\ + \sqrt{\frac{g}{p}}\frac{f_V\rho^2}{\kappa f_\R}\left(\left[\kappa - f_\L\right] u^\mu u_\nu + \frac{f_V}{2} \left[\R u^\mu u_\nu - u^\mu \R\ind{^\alpha_\nu}u_\alpha + 4\R^{\alpha\beta} u_\alpha u_\beta u^\mu u_\nu\right] \right)\\ - \sqrt{\frac{g}{p}} \frac{ff_V}{2f_\R} \rho u^\mu u_\nu\bigg] - \frac{1}{2}\partial_\nu\left(\sqrt{\frac{g}{p}}\left[f-\R f_\R- Vf_V\right]\right)\bigg\} + \rho\C^\alpha_{\beta\nu}u^\beta u_\alpha
\label{eq:deltaPerfectFluid}
\end{multline}
\end{widetext}
[see \prefs{eq:XiPerfectFluid}{eq:PiPerfectFluid} to derive this]. Since $\Delta^\nu_{(P=0)}$ is in general nonzero, the extra force $f^\nu$ is in general nonzero. Hence test particles in $f(\R, \L_m, V)$ gravity do not in general obey the geodesic equation. In other words, test particles traverse geodesics of $g_{\mu\nu}$ if and only if $\Delta^\mu_{(P = 0)} = 0$.

\subsection{The Newtonian limit}

In the weak-field regime we consider the gravitational effect of non-relativistic dust, for which $T_{\mu\nu} = \rho u_\mu u_\nu$ where $u^{\mu} = (\partial_0)^\mu$ is the rest frame four-velocity and $\rho$ is the dust's energy density \cite{Wald}. We shall linearize the $f(\R,\L_m, V)$ equations by keeping terms linear in $\rho$ and in the perturbations introduced below. To facilitate the coming analysis, we adopt the following notation.

Let $\gamma_{\mu\nu}$ and $\widehat{\gamma}_{\mu\nu}$ be smooth two-forms (soon to be perturbations). Further, let $\mathcal{A}$ and $\mathcal{B}$ be mathematical objects composed, in some acceptable fashion, of the objects $\rho, \gamma_{\mu\nu}$, and $\widehat{\gamma}_{\mu\nu}$. Then, by $\mathcal{A} \ll \mathcal{B}$ we shall mean $\mathcal{A}$ is first-order (linear) in at least one of $\rho$, $\gamma_{\mu\nu}$, or $\widehat{\gamma}_{\mu\nu}$, while $\mathcal{B}$ is zeroth-order in all. Moreover, by $\mathcal{A} \cong \mathcal{B}$ we shall mean $\mathcal{A} = \mathcal{B}$ up to at least linear corrections in all $\rho$, $\gamma_{\mu\nu}$, and $\widehat{\gamma}_{\mu\nu}$. Finally, by $\mathcal{A} \sim \mathcal{B}$ we shall mean $\mathcal{A}$ and $\mathcal{B}$ are of the same order in $\rho$, $\gamma_{\mu\nu}$, or $\widehat{\gamma}_{\mu\nu}$, but not necessarily equal (thus $\mathcal{A}\cong \mathcal{B}$ implies $\mathcal{A} \sim \mathcal{B}$).

Consider the metric/auxiliary metric relation posited in \pref{eq:deformationMatrix}. This establishes that any perturbation $\delta p_{\mu\nu}$ upon $p_{\mu\nu}$ satisfies
\begin{equation}
\delta p_{\mu\nu} = g_{\mu\lambda}\delta\Omega\ind{^\lambda_\nu} + \Omega\ind{^\lambda_\nu}\delta g_{\mu\lambda}.
\label{eq:PertubationRequirement}
\end{equation}
Specifically, we shall consider perturbations $\delta p_{\mu\nu} = \widehat{\gamma}_{\mu\nu}$ and $\delta g_{\mu\nu} = \gamma_{\mu\nu}$ upon a Minkowski background $\eta_{\mu\nu}$. Then $p_{\mu\nu} = \eta_{\mu\nu} + \widehat{\gamma}_{\mu\nu}$ and $g_{\mu\nu} = \eta_{\mu\nu} + \gamma_{\mu\nu}$ such that $\widehat{\gamma}_{\mu\nu}, \gamma_{\mu\nu} \ll \eta_{\mu\nu}$. Here $p_{\mu\nu}$ and $g_{\mu\nu}$ are related by \pref{eq:deformationMatrix} and, furthermore, the perturbations $\widehat{\gamma}_{\mu\nu}$ and $\gamma_{\mu\nu}$ satisfy \pref{eq:PertubationRequirement}. Together these imply $\eta_{\mu\lambda} \Omega\ind{^\lambda_\nu} - \eta_{\mu\nu} \ll \eta_{\mu\nu}$, which is possible if and only if $\Omega\ind{^\mu_\nu} \cong \delta\ind{^\mu_\nu} + k\ind{^\mu_\nu}$, where $k\ind{^\mu_\nu} \ll \delta\ind{^\mu_\nu}$. In deriving this, one must also assume that the deformation matrix reacts smoothly to slight perturbations upon the Minkowski background, \emph{i.e.}, that $g_{\mu\lambda}\delta\Omega\ind{^\lambda_\nu} \ll \eta_{\mu\nu}$, which necessitates $\eta_{\mu\lambda}\delta \Omega\ind{^\lambda_\nu} \sim \gamma_{\mu\nu}$ and $\gamma_{\mu\lambda}\delta \Omega\ind{^\lambda_\nu} \cong 0$.

Transcribed to matrix notation, we have $\vec{\Omega} \cong \I + \vec{k}$. Thus, $\vec{\Omega}^{-1} \cong \I - \vec{k}$, to which we shall directly compare \pref{eq:OmegaMatrix}. Since the tensor $T\ind{^\mu_\nu} \rightarrow \vec{g}^{-1}\vec{T}$ is in general different from the identity matrix, it must be that $f_\R \cong \sqrt{\Omega}$ and $f_V \vec{g}^{-1}\vec{T} \cong \sqrt{\Omega}\vec{k}$, where $\sqrt{\Omega} \cong 1 + \frac{1}{2}\Tr(\vec{k})$. But since $\Tr(\vec{k})\vec{k} \cong \vec{0}$, with $\vec{0}$ the zero matrix, we simply have $\vec{k} \cong f_V \vec{g}^{-1}\vec{T}$. Together, these results yield both $p_{\mu\nu}$ and $p^{\mu\nu}$ to the desired first-order precision:
\begin{subequations}
\label{eq:gammaMatrices}
\begin{align}
p_{\mu\nu} & \cong \eta_{\mu\nu} + \gamma_{\mu\nu} + f_V T_{\mu\nu},
\label{eq:psubmunuFirstOrder}\\
p^{\mu\nu} & \cong \eta^{\mu\nu} - \gamma^{\mu\nu} - f_V T^{\mu\nu},
\label{eq:psupmunuFirstOrder}
\end{align}
\end{subequations}
where $T^{\mu\nu} = \eta^{\mu\alpha}\eta^{\nu\beta}T_{\alpha\beta}$ and $\gamma^{\mu\nu} = \eta^{\mu\alpha}\eta^{\nu\beta}\gamma_{\alpha\beta}$. We see from \pref{eq:psubmunuFirstOrder} that
\begin{equation}
\widehat{\gamma}_{\mu\nu} \cong \gamma_{\mu\nu} + f_V T_{\mu\nu}
\label{eq:gammaRelationship}
\end{equation}
and similarly from \pref{eq:psupmunuFirstOrder} that $\widehat{\gamma}^{\mu\nu} \cong \gamma^{\mu\nu} + f_V T^{\mu\nu}$. Hence, the connection coefficients \eref{eq:connectionEquation} are, to linear-order in $\widehat{\gamma}_{\mu\nu}$,
\begin{equation}
\Gamma_{\mu\nu}^\alpha \cong \frac{1}{2} \eta^{\alpha\sigma}\left(\partial_\mu \widehat{\gamma}_{\sigma\nu} + \partial_\nu \widehat{\gamma}_{\mu\sigma} - \partial_\sigma \widehat{\gamma}_{\mu\nu}\right),
\label{eq:approxConnection}
\end{equation}
which yields the Ricci tensor
\begin{equation}
\R_{\mu\nu} \cong \partial^\sigma\partial_{(\nu}\widehat{\gamma}_{\mu)\sigma} - \frac{1}{2}\partial^{\sigma}\partial_\sigma \widehat{\gamma}_{\mu\nu} - \frac{1}{2}\partial_{\mu}\partial_\nu \widehat{\gamma},
\label{eq:approxRicci}
\end{equation}
where $\widehat{\gamma} \equiv \widehat{\gamma}\ind{^\mu_\mu}$. Note that \pref{eq:approxRicci} is entirely of linear-order in $\widehat{\gamma}_{\mu\nu}$. Hence, $V = \R_{\mu\nu}T^{\mu\nu} \cong 0$ since $T^{\mu\nu}$ is linear in $\rho$. Moreover, $f_V\Pi_{\mu\nu} \cong 0$ since $\delta T^{\alpha\beta} / \delta g^{\mu\nu} \sim \rho$ for dust [see \pref{eq:dustPerturbation}] and $\R_{\mu\nu} \sim \widehat{\gamma}_{\mu\nu}$. We shall impose the Lorenz gauge $\partial^\sigma \widehat{\gamma}_{\mu\sigma} = 0$ so that the $f(\R, \L_m, V)$ field equations, to linear-order in $\widehat{\gamma}_{\mu\nu}$ and $\rho$, bear the form
\begin{equation}
-\frac{1}{2}\partial^\sigma\partial_\sigma \widehat{\gamma}_{\mu\nu} - \frac{1}{2}\partial_\mu\partial_\nu \widehat{\gamma} - \frac{1}{2}fg_{\mu\nu} \cong \kappa T_{\mu\nu} - f_{\L}\Xi_{\mu\nu}.
\label{eq:linearF(R,V)}
\end{equation}
To obtain the matter tensor, we necessarily take $\L_m = -\rho$ for the matter Lagrangian of the pressureless dust. Hence, from \pref{eq:dustPerturbation}, $\Xi_{\mu\nu} = \frac{1}{2}\rho(u_\mu u_\nu - \eta_{\mu\nu} - \gamma_{\mu\nu})$. Here $\rho$ is the leading order correction from the matter sector, thus the product $\rho \gamma_{\mu\nu}$ must be regarded as a second-order correction. This implies that, to first-order, $\Xi_{\mu\nu} \cong \frac{1}{2}\rho\left(u_\mu u_\nu - \eta_{\mu\nu}\right)$ and hence $\Xi \equiv \Xi \ind{^\mu_\mu} \cong -\frac{5}{2}\rho$. It follows that the trace of the field equations \eref{eq:FieldEquationOne} is, to first-order, 
\begin{equation}
f_\R \R - 2f \cong \rho\left(\frac{5}{2}f_\L - \kappa \right).
\label{eq:traceFirstOrder}
\end{equation}
Note that $\Pi\ind{^\mu_\mu} \cong 0$ since $\Pi_{\mu\nu} \cong 0$, thus $\Pi\ind{^\mu_\mu}$ is absent \pref{eq:traceFirstOrder} in this approximation. Note also that the already first-order corrections $\rho$ and $\R$ force $f$ to be at least a first-order correction, hence we can rewrite \pref{eq:linearF(R,V)} as
\begin{multline}
-\frac{1}{2}\partial^\sigma\partial_\sigma \widehat{\gamma}_{\mu\nu} - \frac{1}{2}\partial_\mu\partial_\nu \widehat{\gamma} - \frac{1}{2}f\eta_{\mu\nu}\\ \cong \kappa \rho u_\mu u_\nu - \frac{1}{2}\rho f_{\L}\left(u_\mu u_\nu - \eta_{\mu\nu}\right).
\label{eq:linearF(R,V)2}
\end{multline}
The 00 component of this equation encodes the weak-field dynamics in which we are interested. Since the spacetime is assumed static, the time derivatives vanish, leaving the expression
\begin{equation}
-\frac{1}{2}\nabla^2 \widehat{\gamma}_{00} \cong \kappa\rho - \frac{1}{2}f - \rho f_{\L},
\label{eq:almostThere}
\end{equation}
where $\nabla^2$ is the Euclidean space Laplacian. Using \pref{eq:gammaRelationship} and the definition of the Newtonian potential, $\Phi \equiv -\frac{\gamma_{00}}{4}$, we obtain the modified Poisson equation in $f(\R, \L_m, V)$ gravity:
\begin{equation}
\nabla^2 \Phi \cong \frac{1}{2} \kappa \rho - \frac{1}{4}f - \frac{1}{2}\rho f_\L + \frac{1}{4}\nabla^2(f_V \rho).
\label{eq:ModifiedPoisson}
\end{equation}
Since $f$ is implicitly a function of $T_{\mu\nu}$, and hence of $\rho$, the quantity $\frac{1}{2}\kappa\rho - \frac{1}{4}f - \frac{1}{2}\rho f_\L$ acts as a sort of effective density $\frac{1}{2}\kappa\bar{\rho}$. In this notation, \pref{eq:ModifiedPoisson} reads
\begin{equation}
\nabla^2 \Phi \cong \frac{1}{2} \kappa \bar{\rho} + \frac{1}{4}\nabla^2(f_V \rho).
\label{PoissonModifiedTwo}
\end{equation}
This modification to Poisson's equation is formally identical to those in both EiBI and Palatini $f(\R, T)$ gravity (see Refs.~\cite{Jimenez} and \cite{Barrientos}, respectively). Consequently, it is expected that all these theories afford similar non-relativistic phenomenology.

\section{Some Applications}
\label{sec:applications}

The weak-field equations considered above disclosed the relationship between $f(\R, \L_m, V)$ and other theories of gravity in the Newtonian regime. In this section, we derive the field equations governing the response of $f(\R,\L_m, V)$ gravity in other regimes, in particular the electromagnetic and scalar field sectors.

\subsection{Electromagnetic fields}
\label{ssec:EmFields}

We shall consider first the traditional linear electrodynamics (LED) of Maxwell for which the matter Lagrangian is $\L^{(\text{LED})} = -\frac{1}{16\pi}F_{\mu\nu}F^{\mu\nu}$, where $F_{\mu\nu} \equiv \partial_\mu A_\nu - \partial_\nu A_\mu$ is the Faraday tensor. The LED matter tensor follows quickly from its definition \eref{eq:XiTensor}, $\Xi_{\mu\nu}^{(\text{LED})} = -\frac{1}{8\pi} F_{\mu\lambda}F\ind{^\lambda_\nu}$. Alternatively, using \pref{eq:XiAlternative}, 
\begin{equation}
\Xi_{\mu\nu}^{(\text{LED})} = \frac{1}{2}\left(\L^{(\text{LED})}g_{\mu\nu} - T^{(\text{LED})}_{\mu\nu}\right).
\label{eq:XiEM}
\end{equation}
Similarly, from the symmetry of the Ricci tensor and \prefs{eq:PiV2}{eq:XiEM}, the LED matter-curvature tensor turns out to be
\begin{multline}
\Pi^{(\text{LED})}_{\mu\nu} = 2\R\ind{^\lambda_(_\mu}T_{\nu)\lambda}^{(\text{LED})} - \R_{\mu\nu} \L^{(\text{LED})} \\ - \frac{1}{8\pi} \R F_{\mu\lambda}F\ind{^\lambda_\nu} + \frac{1}{4\pi} \R^{\alpha\beta} F_{\mu\beta}F_{\alpha\nu}.
\label{eq:PiEM}
\end{multline}
The LED effective energy-momentum tensor, $\Sigma_{\mu\nu}^{\text{(LED)}}$, then follows trivially from its definition \eref{eq:SigmaTensor}. 

The matter-curvature couplings in \pref{eq:PiEM} are very much unlike, for instance, Palatini $f(\R, T)$ theory, in which $T^{(\text{LED})} = g^{\mu\nu}T_{\mu\nu}^{(\text{LED})} = 0$. Hence, the $f(\R, T)$ models (whether Palatini or not) respond to linear electromagnetic fields as an $f(\R)$ model. This is evidently not the case for the present theory, in which there are nontrivial couplings between curvature and matter terms, all of which have the potential to invite new gravitational electrodynamic behavior. We note that in the Palatini $f(\R, T, V)$ model, all the $f_V$ couplings in \pref{eq:PiEM} persist; therefore, even with a vanishing trace [making the gravitational response a Palatini $f(\R,V)$ theory], there remain new and nontrivial corrections to the linear electrodynamics.

However, it is well known that the linear electrodynamics \emph{in vacuo} are only an approximation to the full electrodynamic theory. General relativity, for example, demands a gravitational coupling between electromagnetic fields, which affords nonlinear electrodynamic behavior. That said, more considerable nonlinearity arises from quantum field effects, such as vacuum polarization \cite{Delphenich}. It is therefore of interest to also derive the $f(\R, \L_m, V)$ field equations associated with a general set of nonlinear electrodynamic (NED) theories. To this end, we set the matter sector action to
\begin{equation}
\S^{(\text{NED})} = \frac{1}{8\pi}\int \d^4x\, \sqrt{-g}\chi(I, J),
\label{eq:NEDaction}
\end{equation}
where $\chi$ is a well-behaved function of the algebraic invariants $I \equiv \frac{1}{2}F_{\mu\nu}F^{\mu\nu}$ and $J \equiv F_{\mu\nu}(\star F)^{\mu\nu}$. Here $(\star F)^{\mu\nu} = \frac{1}{2}(-g)^{-\frac{1}{2}}\epsilon^{\mu\nu \alpha\beta}F_{\alpha\beta}$ is the Hodge dual of the Faraday tensor, with $\epsilon^{\mu\nu\alpha\beta}$ the Levi-Civita symbol. We note that $I$ and $J$ are the unique algebraic invariants constructible from $F_{\mu\nu}$ and $g_{\mu\nu}$ \cite{Landau, Peres}, and also that the choice $\chi(I, J) = -I$ corresponds to the LED theory considered above. 

With $\frac{1}{8\pi}\chi(I, J)$ as the NED matter Lagrangian, and defining $\chi_I \equiv \frac{\partial \chi}{\partial I}$ and $\chi_J \equiv \frac{\partial \chi}{\partial J}$, we find 
\begin{equation}
\Xi_{\mu\nu}^{(\text{NED})} = \frac{1}{8\pi}\left(\chi_I F_{\mu\lambda}F\ind{^\lambda_\nu} + \frac{1}{2} \chi_J J g_{\mu\nu}\right).
\label{eq:XiNED}
\end{equation}
It then follows from \pref{eq:PiV2} and the NED equivalent of \pref{eq:XiEM} that
\begin{widetext}
\begin{multline}
\Pi_{\mu\nu}^{(\text{NED})} = 2\R\ind{^\lambda_(_\mu}T_{\nu)\lambda}^{(\text{NED})} + \frac{1}{8\pi}\left(\chi_J J - \chi\right)\R_{\mu\nu}  + \frac{1}{4\pi}\left(\frac{1}{2}\R \chi_I - \R^{\alpha\beta}F_{\alpha\lambda}F\ind{^\lambda_\beta} \chi_{II} - \frac{1}{2}\R \chi_{JI} J^2 \right)F_{\mu\lambda}F\ind{^\lambda_\nu}\\ -\frac{1}{8\pi} \left(\R^{\alpha\beta} F_{\alpha\lambda}F\ind{^\lambda_\beta} \chi_{IJ} J + \frac{1}{2}\R \chi_{JJ} J^2 \right)g_{\mu\nu} - \frac{1}{4\pi}\R^{\alpha\beta} \chi_I F_{\alpha\mu} F_{\nu\beta}.
\label{eq:PiNED}
\end{multline}
\end{widetext}
As before, $\Sigma_{\mu\nu}^{(\text{NED})}$ then follows from its definition \eref{eq:SigmaTensor}, and $T_{\mu\nu}^{(\text{NED})}$ follows from the NED equivalent of \pref{eq:XiEM}. Note that, as expected, upon fixing $\chi(I,J) = -I$, \pref{eq:PiNED} reduces to \pref{eq:PiEM}. As in the LED case, these field equations have in them nontrivial matter-curvature couplings which again bear new possibilities for NED gravitational dynamics, such as in studies of nonsingular black holes. We also note that these equations again differ drastically in their matter-curvature couplings from the field equations for NED in $f(\R, T)$ gravity (see, \emph{e.g.}, \rref{Barrientos}). This much is evident from the $f_V$ coupling terms, which persist only in the $f(\R, \L_m, V)$ framework.

\subsection{Canonical scalar fields}

Scalar fields comprise another set of generic matter fields for which $f(\R, \L_m, V)$ gravity admits new and nontrivial dynamics. Here we shall consider the effect of a real-valued scalar field $\phi$ in a potential $U(\phi)$, whose Lagrangian density bears the form $\L^{(\phi)} = -\frac{1}{2}\partial_\lambda\phi \partial^\lambda\phi - U(\phi)$. One shall find $\Xi_{\mu\nu}^{(\phi)} = -\frac{1}{2}\partial_\mu\phi\partial_\nu\phi$ and, from \pref{eq:PiV1},
\begin{equation}
\Pi_{\mu\nu}^{(\phi)} = -\frac{1}{2}\R \partial_\mu \phi \partial_\nu \phi + \partial_\lambda \phi \partial_{(\mu}\phi \R\ind{^\lambda_\nu_)} + \R_{\mu\nu} \L^{(\phi)}.
\label{eq:PiScalarField2}
\end{equation}
Hence,
\begin{multline}
\kappa\Sigma_{\mu\nu}^{(\phi)} = \kappa T_{\mu\nu}^{(\phi)} + \frac{1}{2}\left(f_\L + \R f_V\right) \partial_\mu \phi \partial_\nu \phi\\ - f_V\L^{(\phi)}\R_{\mu\nu}- f_V\partial_\lambda \phi \partial_{(\mu}\phi \R\ind{^\lambda_\nu_)}.
\label{eq:effectiveScalar}
\end{multline}
As with the electromagnetic field, specifying particular $f(\R,\L_m, V)$ functions and solving the associated field equations will conceivably yield new non-minimal corrections to ordinary GR problems, which brings about new possibilities. For example, as posited for Palatini $f(\R,T)$ gravity \cite{Barrientos}, free [$U(\phi) = 0$] geonic solutions of the kind in EiBI gravity \cite{Afonso2} are conceivable in the present theory.

\section{Compatibility with EiBI gravity}
\label{sec:EiBI}

In this section we shall investigate the conditions under which the $f(\R, \L_m, V)$ paradigm encapsulates the EiBI theory. We shall denote by $f_{\text{BI}}$ any $f(\R,\L_m, V)$ function that does this. To begin, it is imperative that we be precise with the meaning of ``one gravitational theory corresponding to another.''

Let $\mathscr{A}$ and $\mathscr{B}$ be two Palatini theories of gravity defined on a world-manifold $\M$, and let $\Psi$ be a matter field on $\M$. Further, let $g_{\mu\nu}^{(\mathscr{A})}$ and $g_{\mu\nu}^{(\mathscr{B})}$ be the solutions generated from $\mathscr{A}$ and $\mathscr{B}$, respectively, in response to $\Psi$, and $\nabla^{(\mathscr{A})}$ and $\nabla^{(\mathscr{B})}$ the derivative operators of $\mathscr{A}$ and $\mathscr{B},$ respectively, defined on $\M$. On one hand, we say $\mathscr{A}$ and $\mathscr{B}$ are \emph{equivalent} if, for all $\Psi$, (i) $\nabla^{(\mathscr{A})}_\sigma \xi^\mu = \nabla^{(\mathscr{B})}_\sigma \xi^\mu$ for all vectors $\xi^\mu$ defined on some tangent space in the tangent bundle of $\M$ and (ii) $g_{\mu\nu}^{(\mathscr{A})} = \Theta^2 g_{\mu\nu}^{(\mathscr{B})}$ for some real-valued, smooth conformal factor $\Theta$ defined on $\M$. Evidently, condition (i) ensures that both theories measure the same intrinsic curvature of $\M$, that both have the same notion of transport, and so forth, while condition (ii) establishes that the gravitational dynamics of the two theories be the same (since they afford the same solution, up to a conformal factor, for a given matter sector $\Psi$). On the other hand, we say $\mathscr{A}$ and $\mathscr{B}$ are \emph{circumstantially equivalent} if conditions (i) and (ii) hold only for particular $\Psi$. Indeed, we shall prove in this section that EiBI and $f(\R, \L_m, V)$ are circumstantially equivalent theories of gravity; in particular, that condition (i) shall hold good for all $\Psi$, but that condition (ii) shall hold good only for specific $\Psi$.

The (Palatini) EiBI action bears the form \cite{Jimenez}
\begin{multline}
\hspace{-0.2cm}S_{\text{BI}}[g,\Gamma,\Psi] = \frac{1}{2\kappa\epsilon}\int \d^4x\, \left[\sqrt{\left|g_{\mu\nu} + \epsilon \R_{\mu\nu}(\Gamma)\right|} - \lambda\sqrt{-g}\right]\\ + S_m[g,\Psi],
\label{eq:EiBIAction}
\end{multline}
where $\epsilon$ is a coupling parameter, $\lambda$ is related to the cosmological constant $\Lambda$ by $\lambda = 1 + \epsilon \Lambda$, and the vertical bars denote the absolute value of the determinant. The reader is referred to \rrefs{Banados}{Jimenez} for details on the variation. The field equations are
\begin{subequations}
\label{eq:EiBI}
\begin{align}
q_{\mu\nu} &= g_{\mu\nu} + \epsilon \R_{\mu\nu} \label{eq:EiBI1},\\
\sqrt{-q}q^{\mu\nu} &= \sqrt{-g}\left(\lambda g^{\mu\nu} - \kappa\epsilon T^{\mu\nu}\right),\label{eq:EiBI2}
\end{align}
\end{subequations}
where $q$ is the determinant of the auxiliary metric $q_{\mu\nu}$, and $q^{\mu\nu}$ satisfies both $q^{\mu\lambda}q_{\lambda\nu} = \delta\ind{^\mu_\nu}$ and $\nabla^{(\text{BI})}_\sigma \left(\sqrt{-q}q^{\mu\nu}\right) = 0$, where $\nabla^{(\text{BI})}$ is the derivative operator associated with the Palatini EiBI theory. Hence, $\nabla^{(\text{BI})}_\sigma q_{\mu\nu} = 0$.

We note that $\lambda \neq 0$ (equivalently $\Lambda \neq -\epsilon^{-1}$), for otherwise \pref{eq:EiBI2} implies that \emph{in vacuo} $\sqrt{-q}q^{\mu\nu} = 0$, which is nonsense. Moreover, with $T_{\mu\nu} = 0$ and $\lambda \neq 1$, the solutions from the two theories do not coincide: EiBI affords a de Sitter or anti-de Sitter universe, while $f(\R,\L_m, V)$ outputs Minkowski space. In speaking of a possible equivalence between the theories, it is natural to demand that at least the vacuum solutions correspond. To this end, we shall hereafter fix $\lambda = 1$, making EiBI Minkowskian \emph{in vacuo}. Note that there is no loss of generality in doing this. Should one wish to append a cosmological constant to either theory, they simply do so via the matter sector. We have merely ``tared'' the two theories at the level of their vacuum solutions.

As previously defined, $\nabla^{(p)}$ is the derivative operator associated with the Palatini $f(\R,\L_m, V)$ theory. Thus, for an EiBI/$f(\R, \L_m, V)$ equivalence to exist, condition (i) demands that $\nabla_\sigma^{(\text{BI})} \xi^\mu = \nabla_\sigma^{(p)}\xi^\mu$ for all smooth vectors $\xi^\mu$. This implies, in particular, that
\begin{equation}
\nabla_\sigma^{(\text{BI})}q_{\mu\nu} = \nabla_\sigma^{(p)}q_{\mu\nu} = \nabla_\sigma^{(p)}p_{\mu\nu} = 0.
\label{eq:FundamentalTheorem}
\end{equation}
The connections of both $f(\R,\L_m, V)$ and EiBI gravity are torsion-free. Hence, as required by the fundamental theorem of Riemannian geometry, \pref{eq:FundamentalTheorem} holds good if and only if $p_{\mu\nu} = q_{\mu\nu}$, which is true if and only if $\sqrt{-q}q^{\mu\nu} = \sqrt{-p}p^{\mu\nu}$. Therefore, the definitions \eref{eq:pmunuDefinition} and \eref{eq:EiBI2}, together with condition (ii), \emph{i.e.}, the requisite conformal relationship $g_{\mu\nu}^{(f)} = \Theta^2 g_{\mu\nu}^{(\text{BI})}$ [$g_{\mu\nu}^{(f)}$ being the solution from the $f(\R,\L_m, V)$ theory], imply (with $\lambda = 1$)
\begin{equation}
\left(1 - \Theta^2 f_\R\right)g^{\mu\nu}_{(\text{BI})} - \left(\kappa\epsilon T^{\mu\nu}_{(\text{BI})} + \Theta^4 f_V T^{\mu\nu}_{(f)}\right) = 0,
\label{eq:ConditionThree}
\end{equation}
where $T^{\mu\nu}_{(\text{BI})}$ and $T^{\mu\nu}_{(f)}$ are the energy-momentum tensors of the EiBI and $f(\R, \L_m, V)$ theories, respectively, each raised by their respective metric. We cannot impose \emph{a priori} that these energy-momentum tensors be the same since they are functions of their respective metrics. We can impose, however, that the two parenthetical terms in \pref{eq:ConditionThree} vanish separately. This is necessarily the case if we seek generality in the matter sector, as, for instance, \pref{eq:ConditionThree} holds \emph{in vacuo} if and only if the two parenthetical terms vanish separately. Consequently, $\Theta^2 f_\R = 1$ and $\kappa\epsilon T^{\mu\nu}_{(\text{BI})} = \Theta^4 f_V T^{\mu\nu}_{(f)}$. Differentiating the former with respect to $\R$ demands $f_\R$ to be constant, and hence for the conformal factor $\Theta$ to be constant. Likewise for the latter, where differentiation upon $V$ implies $f_V$ is constant. 

These results indicate two things. First, that $T_{\mu\nu}^{\text{BI}} \propto T_{\mu\nu}^{(f)}$. For equivalence between the two theories to hold, this \emph{constant} proportionality must hold in general, for arbitrary choices of the matter sector. But since the conformal transformation properties of the energy-momentum tensor vary wildly depending on the matter sector, constant proportionality is guaranteed only with exact equality between the metrics, \emph{i.e.}, with $\Theta^2 = 1$ and hence $f_V = -\kappa\epsilon.$ Second, the vanishing of the second derivatives $f_{\R\R}$ and $f_{VV}$ imply that the EiBI/$f(\R,\L_m, V)$ function $f_{\text{BI}}$ is of the form $f_{\text{BI}}(\R, \L_m, V) = f_1(\R) + h(\L_m) + f_2(V)$ for well-behaved functions $f_1, h,$ and $f_2.$ In fact, with the conformal factor set at unity and the energy-momentum tensors identical, we simply have from \pref{eq:ConditionThree} that $f_1(\R) = \R$ and $f_2(V) = -\kappa\epsilon V$. Hence, from  \pref{eq:FieldEquationOne}, the $f_{\text{BI}}$ field equations bear the form
\begin{multline}
\R_{\mu\nu} - \frac{1}{2}\left(\R + h - \kappa\epsilon V\right)g_{\mu\nu} = \kappa T_{\mu\nu} - h' \Xi_{\mu\nu} \\+ 2\kappa\epsilon\R_{\lambda(\mu}T\ind{^\lambda_{\nu)}} + \kappa\epsilon\R^{\alpha\beta}\frac{\delta T_{\alpha\beta}}{\delta g^{\mu\nu}}
\label{eq:f/EiBIFieldEquation}
\end{multline}
where $g_{\mu\nu} = g^{(\text{BI})}_{\mu\nu} = g^{(f)}_{\mu\nu}$, $T_{\mu\nu} = T^{(\text{BI})}_{\mu\nu} = T^{(f)}_{\mu\nu}$, $h' \equiv dh/d\L_m$, and the the identity \eref{eq:PiV2} has been substituted for the matter-curvature tensor. 

We now wish to compare \pref{eq:f/EiBIFieldEquation} with the EiBI field equations \eref{eq:EiBI} so to fix the function $h$ in $f_{\text{BI}}$. However, at the level of the EiBI equations \eref{eq:EiBI}, it is not obvious how to do this. Fortunately, the EiBI equations straightforwardly manipulate into the form \cite{Jimenez}
\begin{equation}
\epsilon \R_{\mu\nu} + \left(1 - \sqrt{\frac{q}{g}}\right) g_{\mu\nu} = \kappa\epsilon T_{\mu\nu} + \kappa\epsilon^2 \R_{\lambda(\mu}T\ind{^\lambda_\nu_)}.
\label{eq:EiBIGeneralization}
\end{equation}
The spacetime is $3 + 1$ dimensional, so $q^{\mu\nu}q_{\mu\nu} = 4$. This allows one to solve explicitly for $\sqrt{q/g}$. \pref{eq:EiBIGeneralization} becomes
\begin{equation}
\R_{\mu\nu} - \frac{1}{4}\left(\R - \kappa T - \kappa\epsilon V\right)g_{\mu\nu} = \kappa T_{\mu\nu} + \kappa\epsilon \R_{\lambda(\mu}T\ind{^\lambda_\nu_)}.
\label{eq:EiBIGeneralizationTwo}
\end{equation}
We wish to investigate what must be true of $h$ and, possibly, $\L_m$ such that the field equations \eref{eq:f/EiBIFieldEquation} and \eref{eq:EiBIGeneralizationTwo} equate to one another. To this end, we set them equal (by solving for $\R_{\mu\nu} - \kappa T_{\mu\nu}$ in both), which, after tracing the two-forms, generates the requisite condition:
\begin{equation}
\R + \kappa\epsilon \R^{\alpha\beta}\frac{\delta T_{\alpha\beta}}{\delta g^{\mu\nu}}g^{\mu\nu} = -2h - \kappa T + h'\Xi.
\label{eq:necessaryConditionOne}
\end{equation}
There are independent ways of satisfying this equation depending on if $\Xi = 0$ or $\Xi \neq 0$. Hence one will have to choose $h$ based on the matter sector under consideration, which demonstrates that $f(\R, \L_m , V)$ gravity is at best circumstantially equivalent to the EiBI framework.

For the former, we assume $\Xi = 0$ identically. Then the matter sector is constant throughout $\M$, implying $\L_m = \Lambda / \kappa$. In this regime, EiBI gravity is known to produce a de Sitter/anti-de Sitter universe equivalent to GR \cite{Banados}. Therefore, $\R = -4\Lambda$, and so, in order for $f(\R, \L_m, V)$ theory to match EiBI theory, one ultimately demands from \pref{eq:necessaryConditionOne} that $h = -2\epsilon \Lambda^2$ so that
\begin{equation}
f_{\text{BI}}(\R, \L_m, V) = \R - 2\epsilon \Lambda^2 - \kappa\epsilon V.
\label{eq:fBIdeSitter}
\end{equation}
This solution implies that $f(\R,\L_m, V)$ gravity can be nontrivially fashioned to have the same de Sitter/anti-de Sitter solutions as both EiBI and GR. It also bespeaks a degeneracy in the $f(\R, \L_m, V)$ framework since the independent (``trivial'') choice $f(\R, \L_m, V) = \R - 2\Lambda$ would just as well deliver the de Sitter/anti-de Sitter solution.

For $\Xi \neq 0$ the process of choosing an $h$ is not as straightforward. We do so in a fashion that shall let us rid constraints on curvature. We note, however, that one could in principle impose constraints on curvature to generate more solutions. In our approach, we shall keep in mind two things. One, $h$ is only a function of the matter Lagrangian density. No terms involving curvature may be appear in its differential equation. Two, not imposing constraints on curvature implies the curvature terms should cancel themselves due to a judicious choice of the matter sector. There is a unique prescription that satisfies these conditions---namely, that $h$ which makes the right side of \pref{eq:necessaryConditionOne} vanish identically, and the corresponding $\L_m$ that makes the left side follow suit.

Demanding the right side of \pref{eq:necessaryConditionOne} to vanish implies the differential equation $-2h - \kappa T + h'\Xi = 0.$ We shall impose $\Xi$ to be nonzero identically, so that one can solve for $h(\L_m)$ explicitly:
\begin{equation}
h(\L_m) = \frac{1}{\omega}\left(C + \kappa\int \d \L_m\, \omega T \right),
\label{eq:hCondition}
\end{equation}
where $C$ is a constant and $\omega$ is an integrating factor given by $\omega(\L_m) = \exp\left(-2 \int \d\L_m\, \Xi^{-1}\right).$ The fact that $C$ is arbitrary implies there is not a unique $f_{\text{BI}}$ when $\Xi \neq 0$, but rather a class of functions for which this particular EiBI/$f(\R,\L_m, V)$ concordance holds good. Now, with the specific choice \eref{eq:hCondition} in hand, we require from \pref{eq:necessaryConditionOne} that $\R + \kappa\epsilon \R^{\alpha\beta} \frac{\delta T_{\alpha\beta}}{\delta g^{\mu\nu}}g^{\mu\nu} = 0.$ This is manifestly true in vacuum (with $\Lambda = 0$). Outside vacuum the condition simplifies by noting $\R_{\mu\nu} \neq 0$ and $\R = \R^{\mu\nu}g_{\mu\nu}$. Hence, by retracting the Ricci tensor, the previous condition necissitates $g_{\mu\nu} + \kappa\epsilon \frac{\delta T_{\mu\nu}}{\delta g^{\alpha\beta}}g^{\alpha\beta} = 0$ for nonzero $\R_{\mu\nu}$. Tracing this expression, and applying algebra upon the identity \eref{eq:EMTensorTwo}, recasts the condition into a form in terms of the matter Lagrangian and trace $T$ of the energy-momentum tensor:
\begin{equation}
g^{\mu\nu}g^{\alpha\beta}\frac{\partial^2 \L_m}{\partial g^{\mu\nu}\partial g^{\alpha\beta}} - 2\L_m + T - \frac{2}{\kappa\epsilon} = 0,
\label{eq:MatterRequirementFour}
\end{equation}
or exclusively in terms of the matter Lagrangian:
\begin{equation}
g^{\mu\nu}g^{\alpha\beta}\frac{\partial^2 \L_m}{\partial g^{\mu\nu}\partial g^{\alpha\beta}} - 2g^{\mu\nu}\frac{\partial \L_m}{\partial g^{\mu\nu}} + 2\L_m - \frac{2}{\kappa\epsilon} = 0.
\label{eq:MatterRequirementFive}
\end{equation}
Consequently, any matter Lagrangian density $\L_m$ satisfying \pref{eq:MatterRequirementFive} and for which $\Xi_{\mu\nu} \neq 0$ will, in the $f(\R,\L_m, V) = \R + h(\L_m) -\kappa\epsilon V$ framework [with $h(\L_m)$ set by \pref{eq:hCondition}], spawn a gravitational response identical to that in EiBI theory. Incidentally, this implies that the non-conservation equation \eref{eq:nonconservationTwo} necessarily vanishes. This follows because EiBI is a minimally coupled theory, as evident from its action \eref{eq:EiBIAction}, hence the metric connection conserves the energy-momentum tensor. Of course, \pref{eq:nonconservationTwo} also vanishes for the nontrivial de Sitter/anti-de Sitter solution \eref{eq:fBIdeSitter}.

\section{Closing remarks}

In this report we have investigated a union of the $f(\R,\L_m)$ and $f(\R, T, \R_{\mu\nu}T^{\mu\nu})$ gravity models in which we allowed arbitrary coupling between the scalar curvature, matter Lagrangian density, and a ``matter-curvature scalar'' $V \equiv \R_{\mu\nu}T^{\mu\nu}$. The model was studied under the Palatini formalism so to generate a bimetric structure comparable to EiBI theory. This implies, in particular, that the independent connection is the Levi-Civita connection of an energy-momentum dependent auxiliary metric that is related to the spacetime metric via a matrix transformation. The equations of motion were derived, and expressed in a manner formally equivalent to $f(\R)$ theories, following the definition of an effective energy-momentum tensor. We briefly described how one obtains the Palatini $f(\R, T, V)$ theory from the present theory, though the exact details of Palatini $f(\R,T,V)$ gravity warrant further investigation.

The field equations impose the non-conservation of the energy-momentum tensor, which gives rise to non-geodesic motion of massive test particles via the appearance of an extra force that will have a nontrivial impact on the physics for compact objects and relativistic stars. In the non-relativistic regime, the dynamics of Palatini $f(\R,\L_m, V)$ gravity are qualitatively similar to the Palatini $f(\R, T)$ and EiBI theories.

With the theory's basic framework established, we introduced the primary elements for some applications. In the case of perfect fluids, the hydrodynamic field equations are nontrivially altered by the non-minimal matter-curvature couplings, even in the non-relativistic regimes. When coupled to electromagnetic fields, either the linear or nonlinear paradigms, the equations have new and nontrivial couplings, and in the case of $f(\R,T,V)$ theory, the electrodynamics reduce to a Palatini $f(\R,V)$ theory due to the vanishing trace. In this realm, the problem of nonsingular black holes can be studied from a separate perspective. Similar remarks apply to scalar fields.

The resemblance to EiBI gravity was then discussed. We showed that $f(\R,\L_m,V)$ gravity is circumstantially equivalent to EiBI, meaning that the two theories have identical spacetime structure and afford identical gravitational dynamics, but only in response to very specific matter fields. 

In summary, the Palatini $f(\R, \L_m, \R_{\mu\nu} T^{\mu\nu})$ gravity theory considered in this work generates a myriad of avenues for future research, and the potential to explore new physics. Further research is expected in this respect, on which we hope to report soon.

\begin{acknowledgments}
The author is indebted to T. Helliwell, V. Sahakian, and B. Shuve for invaluable discussions. % and J. Fox for revisions.
\end{acknowledgments}

\bibliography{EiBIReferences.bib}

\end{document}